\documentclass[preprint,authoryear,12pt]{elsarticle}
\usepackage{epsfig}
\usepackage{amssymb}
\usepackage[ps2pdf,%
a4paper=true,%
breaklinks=true,%
colorlinks=true,%
pdfauthor={First Author et al.},%
pdftitle={Template for manuscripts in Advances in Space Research}%
]{hyperref}

\journal{Advances in Space Research}
 
\begin{document}


\begin{frontmatter}

\title{Comptonizing Efficiencies of IGR 17091-3624 and its similarity to GRS 1915+105}

\author{Partha Sarathi Pal\corref{cor}\fnref{footnote1}}
\address{S. N. Bose National Centre For Basic Sciences, Kolkata.}
\cortext[cor]{Corresponding author}
\fntext[footnote1]{S. N. Bose National Centre For Basic Sciences, Kolkata.
JD Block, Salt Lake, Kolkata - 700098, India.
Tel: +91-33-2335-5706/7/8 Extn: 374,
Fax: +91-33-2335-9176.
}
\ead{parthasarathi.pal@gmail.com, partha.sarathi@boson.bose.res.in}

\author{Sandip K. Chakrabarti\fnref{footnote1}}
\ead{chakraba@bose.res.in}

\begin{abstract}

Variability classes in the enigmatic black hole candidate GRS 1915+105 are
known to be correlated with the variation of the Comptonizing Efficiency (CE) which is 
defined to be the ratio between the number of power-law (hard) photons 
and seed (soft) photons injected into the Compton cloud. Similarities of 
light curves of several variability classes of GRS 1915+105 and IGR 17091-3624, 
some of which are already reported in the literature, motivated us to compute CE for 
IGR 17091-3624 as well. We find that they are similar to what were reported earlier 
for GRS 1915+105, even though masses of these objects could be different.
The reason is that the both the sizes of the sources of the seed photons and of 
the Comptonizing corona scale in the same way as the mass of the black hole. 
This indicates that characterization 
of variability classes based on CE is likely to be black hole mass independent, in general.
\end{abstract}

\begin{keyword}
{Black Holes, Accretion disk, X-rays, Radiation mechanism.}
\end{keyword}

\end{frontmatter}

\parindent=0.5 cm

\section{Introduction}

It has been shown recently that the variability classes of GRS 1915+105 could be characterized by 
Comptonizing Efficiency (CE) which is defined by the ratio of the power-law photons and injected 
seed photons averaged over the class duration (\citet{p13} and references therein). 
Since the number of photons producing the power-law component is the same as the intercepted 
seed photons by the Comptonizing region, in reality, CE measures the geometry (more accurately, the
optical depth) of the 'Compton cloud'. Not surprisingly, it was found that CE is very small 
for softer classes where the 
Compton cloud collapses as it cooled down and larger for harder classes where the Compton 
cloud is large. Earlier, for GRS 1915+105, these classes were characterized by conventional photon number 
ratios (so-called hardness ratios) in specific energy bands \citep{paul97, paul98, yad99, mun99, b00}.  
Such ratios, and especially color-color diagrams drawn based on these ratios, appear to be different when 
a black hole of another mass is chosen since the definition of soft photon energy and hard photon energy 
depends on the mass. Indeed, one will have to fine tune energy bands suitably
to obtain color-color diagrams of similar appearance for the same type of variability class. 
However, our motivation stems from the natural outcome of the two component 
advective flow (TCAF) paradigm (\citet{ct95}, hereafter CT95).  
Here, both the sources of soft photons (Keplerian disk) and Comptonized photons 
(Post-shock region of the advective flow component) are different regions of the TCAF 
and both of them scale with the mass of the black hole. 
Therefore our classification is expected to be mass independent. 
Thus CE, which is something similar to hardness ratio averaged over a short time scale  
compared to a class duration, can be used for defining variability classes. If CE remains similar 
for two objects having similar light curves, it also implies that they have {\it similar} 
spectral behavior and thus the classification is not superficial. 
In \citet{p13}, we prescribed a sequence in which all the variability classes of 
GRS 1915+105 are expected to be manifested. Preliminary indications show \citep{p11}
that indeed, observed variability class transitions follow our sequence. 
Our geometry based discussion presented above would therefore imply that for any other object, 
the sequence must also be the same. This makes our
procedure to be a powerful tool not only to replace conventional hardness ratio plots by CE plots, 
but also has a predictability in terms of how the light curves and spectra are likely to appear in future.

IGR J17091-3624 was first observed by an INTEGRAL Galactic Center Deep Exposure Coverage \citep{kul03}.  
Later, the position was confirmed through Swift observation leaving two nearby blended candidates 
\citep{ken07, cha08}. In early 2011, Swift-BAT detected an outburst from IGR J17091-3624 \citep{krim11}.
Optical/IR observations of the compact object was performed before and during the outburst 
\citep{tor11}. In this outburst, radio emission was detected which confirmed the 
presence of an accreting black hole candidate in its low-hard state \citep{cor11, rod11}. 
Rossi X-ray Timing Explorer (RXTE) detected low-frequency quasi-periodic oscillations \citep{rod11} and  
milli-Hertz and high frequency QPOs \citep{al11d, al12}. Some variability classes of the light curve that 
were seen in GRS 1915+105 were also observed for IGR 17091-3624 \citep{al11a, al11b, al11c, pa12}. 
As IGR J17091-3624 has a lower luminosity than GRS 1915+105 \citep{rod11}, it is assumed that 
IGR J17091-3624 either consists of a very low mass black hole \citep{al11c} and/or it is located 
far away in the Galaxy with a distance of around $17 - 20$ kpc \citep{rod11, al11c}. 

In the present paper, we show that IGR J17091-3624 has many apparently visually `similar looking' 
variability classes (more than what is reported in the literature so far) 
as those of GRS 1915+105, though the time scales are different. 
Most surprisingly, we show that `similar looking' variability 
classes of the two objects yield {\it similar Comptonizing Efficiencies or CEs}
despite the fact that their masses could be totally different. 
Thus we can characterize each variability class by a unique CE value, independent 
of the mass of the black hole. Being similar CEs, the sequence in which hitherto 
unobserved class transitions are expected to take place for IGR J17091-3624 
also becomes predictable. Since CE carries the information about geometry of the 
Compton cloud, we conclude that the Compton cloud geometry in IGR J17091-3624 also evolves
in the same way as that in GRS 1915+105 \citep{p11, p13, p14}. 

In the next Section, we present our analysis procedure and computation of Comptonizing 
Efficiency or CE for the IGR 17091-3624. We then present results of analysis of IGR 17091-3624. 
We compare CE values of IGR 17091-3624 with CE values of GRS 1915+105. Finally, we draw our conclusions.

\section{Analysis \& Calculation of CE}

In order to obtain spectrum in wider energy range, 
we have analysed data in the common good time intervals of  RXTE-PCA and Swift-XRT 
in which IGR 17091-3624 was observed on days mentioned in Table.~\ref{tbl1}. 

\begin{table}
\begin{center}
\caption{Simultaneous Observation IDs along with common good time intervals 
of RXTE-PCA and Swift-XRT data of IGR 17091-3624 analysed in this paper. \label{tbl1}}
\begin{tabular}{cccc}
\hline
\hline
Observation & Swift-XRT & RXTE-PCA & Common GTI\\
date &  data id & data id  & sec \\
\hline
06/04/2011 &  00031921040 & 96420-01-06-03 & 461  \\
10/05/2011 &  00031921049 & 96420-01-11-03  &  880 \\
20/05/2011 &  00031921053 & 96420-01-13-00  &1248 \\
17/07/2011 &  00035096020 & 96420-01-21-02 & 411 \\
29/07/2011 &  00035096025 & 96420-01-23-00  & 1057 \\
31/07/2011 &  00035096026 & 96420-01-23-02  & 801 \\
02/08/2011 &  00035096027 & 96420-01-23-04  & 454  \\
04/08/2011 &  00035096028 & 96420-01-23-06 & 875  \\  
24/09/2011 &  00035096044 & 96420-01-31-01 & 2559 \\
22/10/2011 &  00035096057 & 96420-01-35-00 & 552 \\
26/10/2011 &  00035096059 & 96420-01-35-03 & 918 \\
\hline
\end{tabular}
\end{center}
\end{table}

\begin{figure}
\begin{center}
\includegraphics[width=10.0truecm,angle=0]{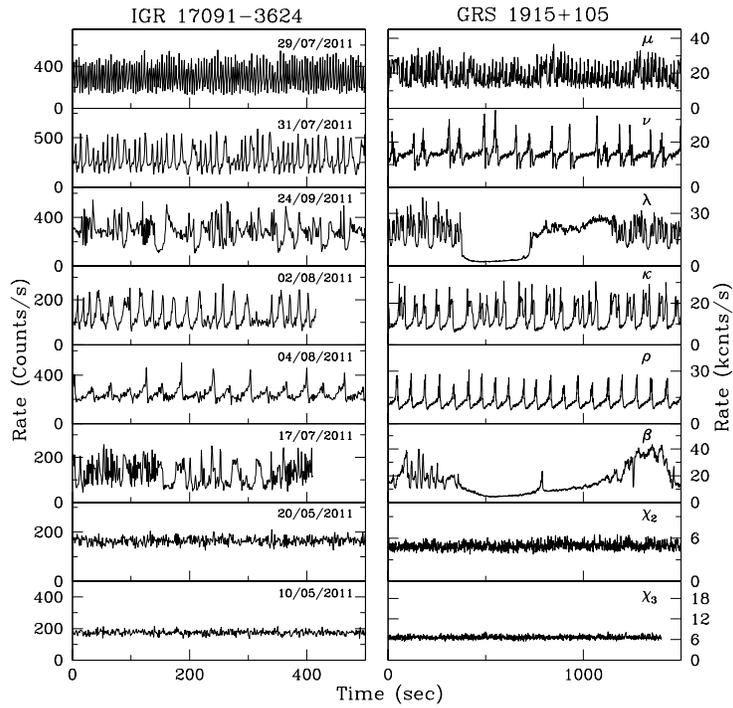}
\caption{Comparison of the 2.0 - 40.0 keV 1.0 sec time bin RXTE-PCA
light curves of IGR 17091-3624 (left) and GRS 1915+105 (right) to show their similarities.
IGR 17091-3624 light curves clearly resemble with $\mu, \nu, \lambda, \kappa,
\rho, \beta, \chi_2$ and $\chi_3$ classes of GRS 1915+105. \label{fig1}}
\end{center}
\end{figure}

We plot $2.0 - 40.0$ keV RXTE-PCA light curves of 1.0 sec time bin of both IGR 17091-3624 
and GRS 1915+105 in panels of Fig.~\ref{fig1}. Several workers reported that IGR 17091-3624 sometimes 
shows variabilities similar to GRS 1915+105 \citep{al11a, al11b, al11c, pa12}. We show these 
and some more variability classes discovered by us in Fig.~\ref{fig1} by plotting them 
side by side for comparison. Some of these (such as, $\beta$) may not look similar 
superficially as a whole. This is because duration of a dwell was not as large as should have been for 
IGR 17091-3624. If sufficient time was given to observe IGR 17091-3624, or time was scaled appropriately, 
the shapes would exhibit more resemblances.

\subsection{Spectral Analysis}

We analyze RXTE-PCA and Swift-XRT data given in Table.~\ref{tbl1}. 
We chose data sets by common MJDs of the outbursts observation, obtained from 
the NASA archive for both instruments. This allows us to study spectral properties
in the wider range of energy. These data are downloaded from NASA Archive at HEASARC.
RXTE and Swift data are reduced with HEASOFTv6.12 software. 

For RXTE data reduction, we exclude data collected for elevation angles less than 
$10^{\circ}$, for offset greater than $0.02^{\circ}$ and those acquired during the 
South Atlantic Anomaly (SAA) passage. We selected RXTE PCU2 data as it was active 
during the time of observation. 
The RXTE-PCA spectra are extracted using ``standard2" mode data which 
have $16$ sec time resolution and energy selection from $3.0$ keV to $25$ keV. 

For Swift data reduction, the level 2 cleaned event files of SWIFT-XRT 
are obtained from window grade 0 events of timing (WT) mode data with {\it xrtpipeline}. 
The spectra are extracted from a 40 $\times$ 10 pixel region in the best source position. 
The background is estimated from an off-axis region of the same size.
The ancillary response files (arfs) are extracted with {\it xrtmkarf}. The
WT redistribution matrix file (rmf) version (v.12) is used in the spectral fits.
The Swift-XRT spectra are extracted with $16$ sec time resolution and energy
selection from $0.5$ keV to $10$ keV. 

The combined $0.5$ keV to $25$ keV spectrum, consists of $0.5$ keV to $10$ keV Swift-XRT spectrum 
and $3.0$ keV to $25$ keV RXTE-PCA spectrum of the same good time interval and the same spectral 
resolution, are analysed using software package XSPEC 12.8.2. Here, each combined spectrum 
is averaged for $16$ sec.   

All spectra are fitted with {\it diskbb} and {\it power-law} model along with 
hydrogen column density for absorption due to interstellar medium 
nH $1.1 \times 10^{22} cm^{-2}$ \citep{k11}.
We have multiplied a constant value along with the model components to compensate for the 
difference of normalizations between SWIFT-XRT and RXTE-PCA. 
During fitting of all spectra we used the technique introduced by \citet{s99b,p11,p13} 
to obtain spectral parameters to calculate number of seed photons and Comptonized photons (described below). 
We consider error-bars at 90\% confidence level in each case under consideration. 


\subsection{Calculation of Comptonizing Efficiency (CE)}

The fitting parameters are used to calculate the black body photons from the Keplerian disk and
the power-law photons from the hot electron cloud. The number of black body photons  
are obtained from the fitted parameters of the multi-color disk black body model \citep{maki86}. 
This is given by,
\begin{equation}
f(E)=\frac{8\pi}{3} r_{in}^2 \cos{i} \int_{T_{out}}^{T_{in}} (T/T_{in})^{-11/3}B(E,T)\,dT/T_{in},
\end{equation}
where, $B(E,T)=\frac{E^3}{(\exp{E/T}-1)}$ and $r_{in}$ can be calculated from,
\begin{equation}
K=(r_{in}/(D/10kpc))^2 \cos{i},
\end{equation}
where, $K$ is the normalization of the blackbody spectrum obtained after fitting, $r_{in}$ is the
inner radius of the accretion disk in $km$, $T_{in}$ is the temperature at $r_{in}$ in keV, $D$ is
the distance of the compact object in kpc and $i$ is the inclination angle of the accretion
disk. Here, both the energy and the temperature are in keV. The black body flux, $f(E)$ in
photons/s/keV is integrated between $0.1$ keV to the maximum energy $dbb_e$. $dbb_e$
is the highest energy up to which the spectrum is fitted with diskbb model alone 
\citep{s99b,p11,p13}. This gives us $N_{BB}$, the rate at which black body photons are emitted.   

\begin{figure}
\begin{center}
\includegraphics[width=7.5truecm,angle=270]{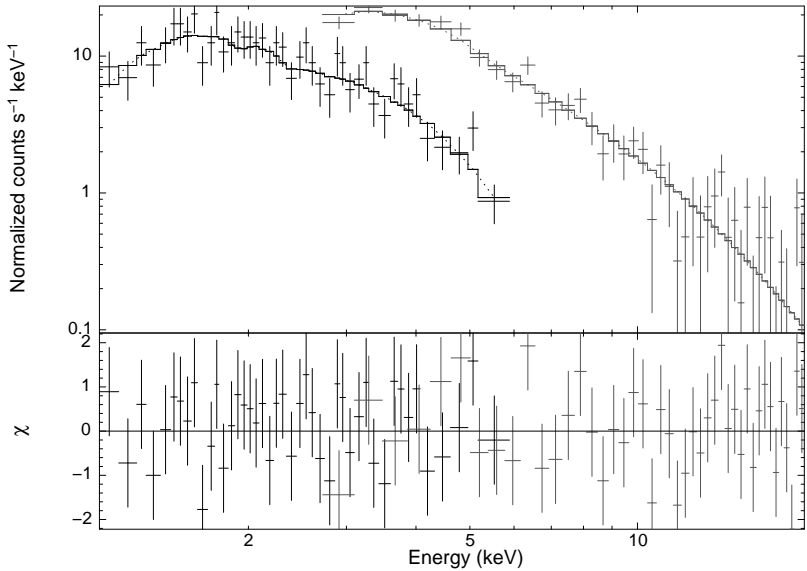}
\caption{A simultaneous 16 sec sample of XRT and PCA fitted spectra during 04/08/2011. 
The black curves represents the SWIFT-XRT data and gray curves represent RXTE-PCA data. 
The spectra are fitted with {\it diskbb} and {\it power-law} components. \label{fig2}}
\end{center}
\end{figure}

The Comptonized photons $N_{PL}$ that are produced due to inverse-Comptonization of the soft 
black body photons by `hot' electrons in the Compton cloud are calculated by fitting with 
the power-law given below,
\begin{equation}
P(E)=N E^{-\alpha},
\end{equation}
where, $\alpha$ is the power-law index and $N$ is the total $photons/s/cm^2/keV$ at $1$keV.
It is reported \citet{tit94} that the Comptonization spectrum will have a peak at around
$3 \times T_{in}$. The power-law equation is integrated from $3 \times T_{in}$ to $40$keV to
obtain the rate of emitted Comptonized photons. We have chosen the upper energy limit 
for calculation of Comptonized photons to be $40$ keV as this is the upper energy limit of the PCA
detector. We do not anticipate that our results would change significantly even when 
higher energy photons are included since the number of photons decreases rapidly with energy. 
In Table.~\ref{tbl2} we give parameters of spectral analysis of XRT and PCA spectrum. 
Fig.~\ref{fig2} shows a fitted 16 sec sample
of simultaneous SWIFT-XRT and RXTE-PCA spectra for IGR 17091-3624 during 04/08/2011. 
The spectra are fitted with diskbb and power-law components. The photon numbers and CE 
are calculated from the fitting parameters. 
The diskbb spectrum is simulated with $T_{in} = 1.31^{+0.07}_{-0.07}$ keV. 
The calculated number of blackbody photons from 0.1-6.0 keV is $18.42^{+3.46}_{-2.87}$ kcnts/s.
Then spectra fitted with {\it diskbb} and {\it power-law} while diskbb components remained 
frozen to the values obtained from previous fitting of spectra with diskbb model only. 
The power-law spectrum is simulated with power-law index=$1.19^{+0.39}_{-0.76}$. 
The calculated number of Comptonized photons between $3.93-40.0$ keV is 
$0.04^{+0.005}_{-0.006}$ kcnts/s. 
This means that only $0.21^{+0.07}_{-0.08}$\% blackbody photons are actually Comptonized. 
For the intermediate states we have computed twice, one for burst-off (low count) state 
which is harder (denoted by 'h') and the other for the burst-on state (high count) which is 
softer (denoted by 's'). Details of these states are presented in \citet{p13} and references therein. 
   
\begin{table}
\scriptsize{
\addtolength{\tabcolsep}{-2.75pt}
\begin{center}
\caption{Parameters for the spectral fits of simultaneous 16s bin of PCA and XRT 
spectra with {\it diskbb} plus {\it power-law} models for all observation dates. 
$h$ and $s$ are sample parameters for harder and softer states in a class with
intermediate states. $T_{in}$ is the black body temperature obtained from fitting. 
$dbb_e$ is the upper limit of the disk blackbody spectrum. $\tilde{{\chi_\dag}^2}$ 
represents the reduced $\chi^2$ value of 0.9-$dbb_e$ keV spectra fitted with diskbb model only. 
Soft photon rate includes blackbody photons in the $0.1-dbb_e$ keV. 'power-law'
is the power-law index $\alpha$ obtained from fitting. Hard photon rate includes 
Comptonized photons emitted in $3 \times T_{in} - 40$ keV. CE is the Comptonized Efficiency. 
$\tilde{{\chi_\ddag}^2}$ is reduced $\chi^2$ value of $0.9-25.0$ keV. The sequence is according to the 
increased CE averaged over a variability class (same as in Fig.~\ref{fig5} below).
\label{tbl2}}
\begin{tabular}{|c|c|c|c|c|c|c|c|c|c|}
\hline
\multicolumn{2}{|c|}{} & $T_{in}$ & $dbb_e$ & $\tilde{\chi_\dag}^2$ & Soft & Power law &  Hard & CE  & $\tilde{\chi_\ddag}^2$ \\
\multicolumn{2}{|c|}{Date} &  &  & & Photon & index &  Photon &  &  \\
\multicolumn{2}{|c|}{} & (keV) & (keV) & (dofs) & (kphtns/s) &   & (kphtns/s) & (\%) & (dofs) \\
\hline
29/07/2011 &-&$1.66^{+0.09}_{-0.08}$& 10 & 0.94(65) & $16.89^{+2.96}_{-2.47}$ & $1.95^{+0.52}_{-0.50}$ & $0.2^{+0.02}_{-0.01}$ & $0.1^{+0.02}_{-0.03}$ & 0.88(114)\\ 
\hline
26/10/2011 &-&$1.84^{+0.12}_{-0.11}$& 10 & 1.34(64) &$14.48^{+3.09}_{-2.47}$ & $2.50^{+0.50}_{-0.52}$ &$0.01^{+0.005}_{-0.002}$ & $0.07^{+0.04}_{-0.03}$ & 1.08(114)\\ 
\hline
22/10/2011 &-& $1.97^{+0.13}_{-0.12}$& 10 & 1.24(62) & $15.02^{+3.41}_{-2.68}$ & $2.15^{+0.40}_{-0.40}$ &$0.02^{+0.006}_{-0.005}$ & $0.12^{+0.07}_{-0.02}$ & 1.06(112) \\ 
\hline
31/07/2011 &h& $1.45^{+0.08}_{-0.07}$& 10 & 1.09(65) & $19.45^{+3.44}_{-2.89}$ & $1.86^{+0.50}_{-0.50}$ &$0.02^{+0.003}_{-0.004}$ & $0.10^{+0.03}_{-0.04}$ & 0.95(114) \\ 
\cline{2-10}
          &s& $1.57^{+0.09}_{-0.08}$& 10 & 0.86(57) & $16.19^{+3.00}_{-2.48}$ & $2.12^{+0.57}_{-0.56}$ & $0.01^{+0.003}_{-0.002}$ & $0.06^{+0.02}_{-0.02}$ & 0.78(106) \\ 
\hline
24/09/2011 &-& $1.48^{+0.08}_{-0.08}$& 10 & 1.35(73) & $20.39^{+3.73}_{-3.10}$ & $2.07^{+0.37}_{-0.72}$ & $0.03^{+0.008}_{-0.009}$ & $0.12^{+0.02}_{-0.06}$ & 1.05(123) \\ 
\hline
02/08/2011 &h& $1.53^{+0.1}_{-0.1}$& 10 & 1.59(54) & $13.82^{+3.34}_{-2.61}$ & $1.61^{+0.25}_{-0.31}$ &$0.03^{+0.002}_{-0.002}$ & $0.23^{+0.07}_{-0.06}$ & 1.02(103) \\ 
\cline{2-10}
          &s& $1.56^{+0.08}_{-0.07}$& 10 & 1.28(71) & $19.20^{+3.29}_{-2.75}$ & $2.30^{+0.46}_{-0.78}$ &$0.02^{+0.004}_{-0.008}$ & $0.10^{+0.04}_{-0.06}$ & 1.06(120) \\ 
\hline
04/08/2011 &h& $1.31^{+0.07}_{-0.07}$& 6.0 & 1.09(40) & $18.42^{+3.46}_{-2.87}$ & $1.19^{+0.39}_{-0.76}$ &$0.04^{+0.005}_{-0.006}$ & $0.21^{+0.07}_{-0.08}$ & 0.97(112) \\ 
\cline{2-10}
          &s& $1.53^{+0.10}_{-0.09}$& 9.0 & 0.99(53) & $13.95^{+3.20}_{-2.55}$ & $2.00^{+0.03}_{-0.02}$ &$0.01^{+0.006}_{-0.006}$ & $0.07^{+0.03}_{-0.06}$ & 0.95(102) \\ 
\hline
06/04/2011 &h& $1.16^{+0.10}_{-0.09}$& 3.0 & 0.87(28) & $16.48^{+4.73}_{-3.60}$ & $1.49^{+0.63}_{-0.85}$ & $0.03^{+0.016}_{-0.014}$ & $0.20^{+0.15}_{-0.13}$ & 0.75(111) \\ 
\cline{2-10}
          &s& $1.40^{+0.20}_{-0.11}$& 6.0 & 1.05(40) & $17.46^{+4.04}_{-3.78}$ & $2.17^{+0.45}_{-0.75}$ &$0.01^{+0.004}_{-0.006}$ & $0.09^{+0.06}_{-0.05}$ & 0.86(124) \\ 
\hline
17/07/2011 &h& $1.14^{+0.11}_{-0.09}$& 6.0 & 1.66(23) & $84.59^{+15.09}_{-19.90}$ & $2.46^{+0.78}_{-0.53}$ & $0.27^{+0.050}_{-0.002}$ & $0.32^{+0.17}_{-0.08}$ & 0.79(90) \\ 
\cline{2-10}
          &s& $1.87^{+0.19}_{-0.18}$& 10 & 0.79(90) & $61.30^{+21.09}_{-16.55}$ & $2.98^{+0.32}_{-0.27}$ &$0.09^{+0.020}_{-0.026}$ & $0.15^{+0.04}_{-0.04}$ & 0.82(82) \\ 
\hline
20/05/2011 &-&$1.12^{+0.21}_{-0.17}$& 6.5 & 1.3(37) & $13.73^{+5.14}_{-3.54}$ & $2.19^{+0.24}_{-0.30}$ &$0.08^{+0.01}_{-0.01}$ & $0.58^{+0.29}_{-0.24}$ & 0.95(95)  \\ 
\hline
10/05/2011 &-&$1.11~^{+0.26}_{-0.17}$& 3.1 & 1.0(21) &$7.52^{+6.45}_{-2.96}$ & $2.02^{+0.33}_{-0.63}$ &$0.06^{+0.08}_{-0.02}$ & $0.77^{+0.37}_{-0.38}$ & 0.93(96)  \\ 
\hline
\end{tabular}
\end{center}}
\end{table}




At any given instant, the ratio $N_{PL}$/$N_{BB}$ is the Comptonizing Efficiency (CE). 
This is similar to a `hardness ratio' though the energy ranges of the hard and the soft photons are 
automatically determined by the fitting process as described above. In TCAF paradigm (CT95),
power-law photons originate through Comptonization of the post-shock region of the low-angular momentum
advective component and seed photons are from the Keplerian component. However, sizes of both
the components scale with the mass of the black hole and thus we expect CE to be mass independent.
Thus using CE in place of conventional hardness ratio would would be more appropriate in 
classification of various light curves and variability types. 

\section{Results}

On 06/04/2011, the PCA count rate varies from $50$ to $200$ counts/s. The light curve
matches with a part of $\beta$ class of GRS 1915+105. Here, the value of CE 
varies from 0.08\% to 0.45\% during softer and harder states respectively. 

On 10/05/2011, the rate is steady at around $170$ counts/s as in the $\chi_3$ class.  
The object is in a hard state. A QPO at around $5$ Hz is observed during this observation. 
Here the amount of CE is around $0.8\%$. CE remains more of less steady for the entire 
observation dwell.

On 20/05/2011, the rate is  steady at around $170$ counts/s and the light curve 
resembles that of the $\chi_2$ class of GRS 1915+105. The object is in a hard state. 
A QPO at around $4$ Hz is observed. Here CE remains steady around $0.4\%$ throughout 
the entire observation dwell.

On 17/07/2011, the rates vary in the range of $50-200$ counts/s. This is a variable
intermediate class, most likely a part of the $\beta$ type class of GRS 1915+105,
and no fixed periodicity is observed during this observation. 
Here, CE is varying from 0.04\% to 0.4\% during its softer and harder states 
respectively. 

On 29/07/2011, the rate is around $300$ counts/s. During this observation
no quasi periodic oscillation (QPO) is observed. Average CE is $\sim 0.08 \%$. 
This class looks very similar to the $\mu$ class of GRS 1915+105.

On 31/07/2011, the rate changes to around $150-350$ counts/s. This light 
curve looks similar to $\nu$ type variability class of GRS 1915+105. 
Here the flickering takes place around each $10$ sec interval. Average 
CE varies around $\sim 0.11\%$. No distinct QPO is observed during this class.  

On 02/08/2011, the rate remains in the range of $\sim 100-200$ counts/s. The light curve
shows a similar variability as in the $\kappa$ class of GRS 1915+105. 
Some periodicity of $\sim 10-15$ sec is observed also. The average CE is 
around $\sim 0.16\%$ during this observation. 

\begin{figure}
\includegraphics[angle=-90,scale=0.27]{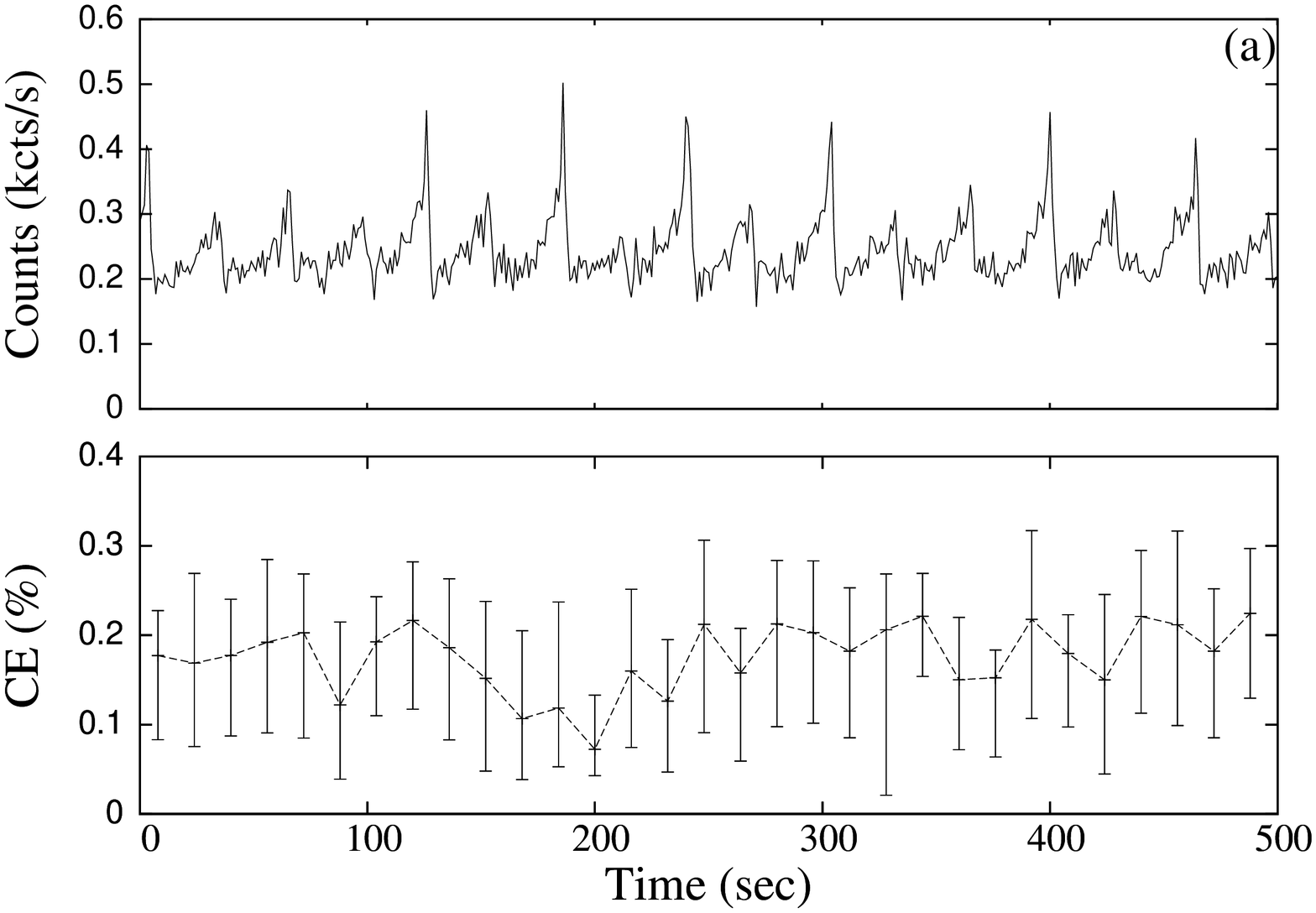}
\includegraphics[angle=-90,scale=0.27]{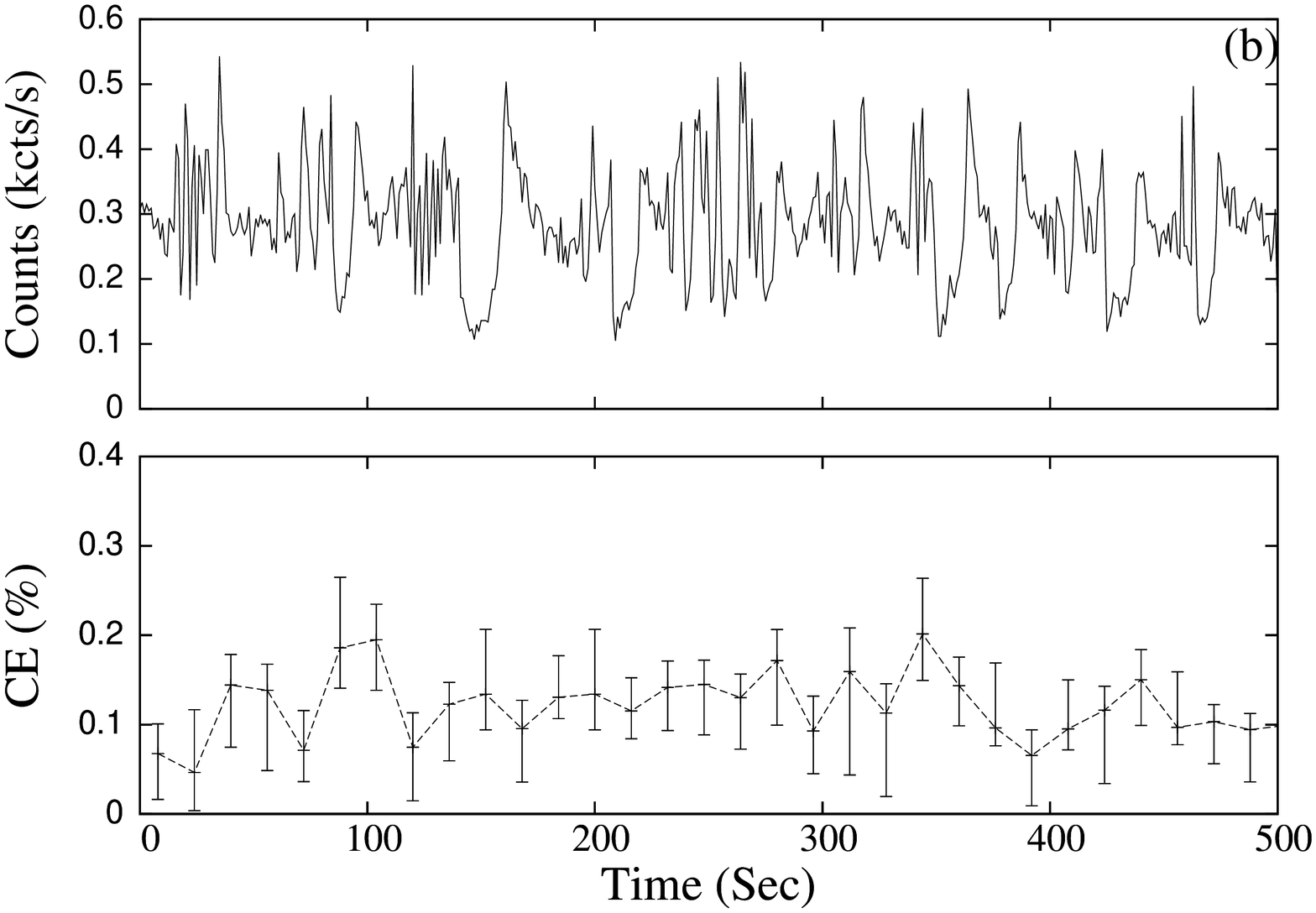}
\caption{(a) Top panel shows 2.0 - 40 keV and 1.0 sec time bin RXTE-PCA light curve of
IGR 17091-3624 of 04/08/2011 observation. Bottom panel shows variation of average 
CE with time as obtained from 16s binned data for this observation. 
(b) Top panel shows 2.0 - 40 keV and 1.0 sec time bin RXTE-PCA light curve of
IGR 17091-3624 of 24/09/2011 observation. Bottom panel shows variation of average 
CE with time as obtained from 16s binned data for this observation. 
\label{fig3}}
\end{figure}

On 04/08/2011, the rate varies in the range of $200$ to $350$ counts/s. The light curve is 
variable as in $\rho$ class of GRS 1915+105. The interval of each 'heartbeat'
is around $25$ sec. During this observation, CE is varying around $0.07\% - 0.2\%$ 
during their softer and harder states respectively. We plot CE variation for 
the 04/08/2011 observation of IGR 17091-3624 data as in Fig.~\ref{fig3}(a). 

In Fig.~\ref{fig3}(a), we plot CE as a function of time. Since the spectra are necessarily 
from a $16$s bin, CE does not show the same rise and fall as that of the light curve because 
of averaging out effects arising out of the minimum bin time constraints. 

On 24/09/2011, the rate varies between $100-500$ counts/s. The light curve shows a variable 
shape but it does not show any regular periodicity. This can be treated to be a variable intermediate 
class, likely to be  a part of $\lambda$ class of GRS 1915+105. 
During this observation, CE is found to vary between 0.04\% and 0.2\% in the burst-off and
burst-on states respectively. 
We plot CE variation for this observation of IGR 17091-3624 data as in Fig.~\ref{fig3}(b). 
In Fig.~\ref{fig3}(b), we plot CE as a function of time. Since the spectra are necessarily 
from a $16$s bin, CE does not show the same rise and fall as that of the light curve because 
of averaging out effects arising out of the minimum bin time constraints. 

On 22/10/2011, the rate is around $140$ counts/s. In this observation also, no clear 
periodicity is observed. On an average, CE is $\sim 0.10\%$ similar to that in the 
$\mu$ class of GRS 1915+105. 

Finally, on 26/10/2011, the rate varied between $300-350$ counts/s. This variable class also 
looks like the $\mu$ class of GRS 1915+105 with no clear periodicity. In this observation, CE 
varies in the range of 0.04\% to 0.1\%. 

\section{Discussion \& Conclusion}

We compute CE for each observation. Note that in no occasion, 
transition from one variability class to another was observed during a single observation dwell. 
In future, with continuous monitoring such transitions are expected to be seen. 

\begin{figure}
\includegraphics[angle=0,scale=0.68]{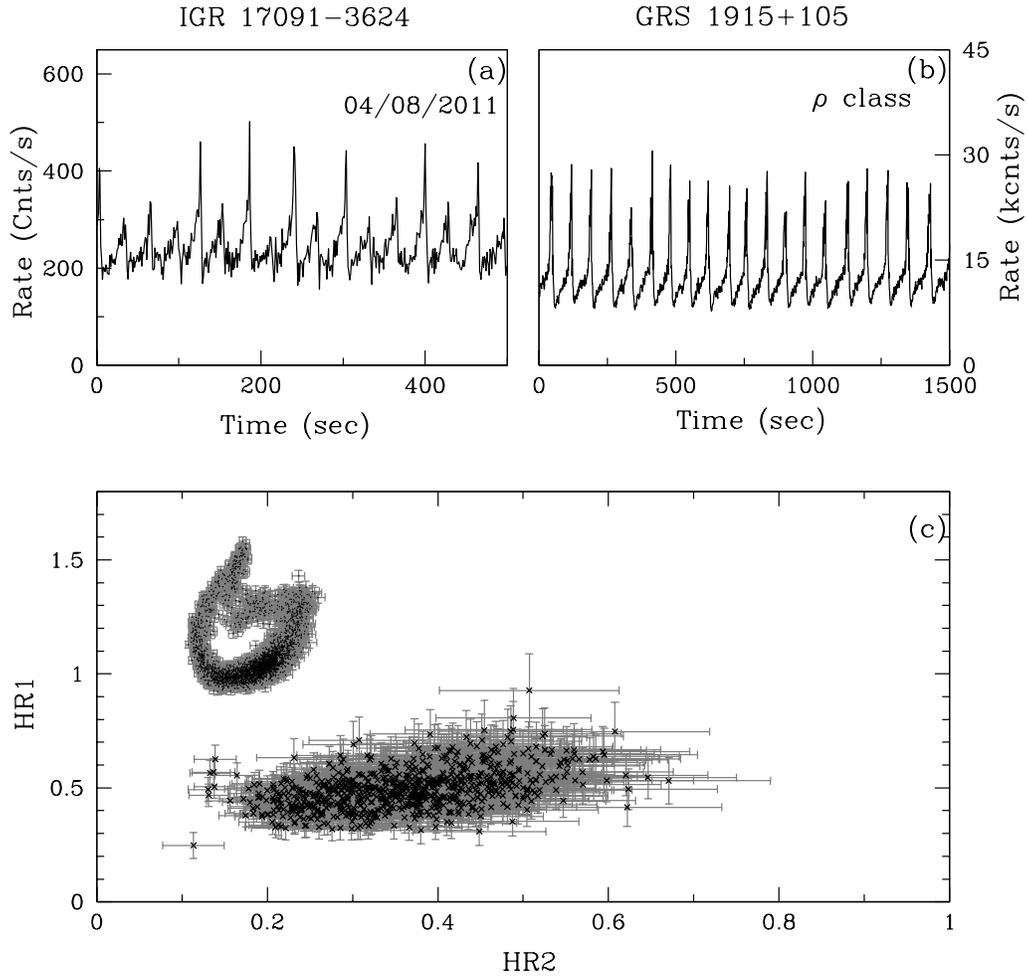}
\caption{(a) 2 - 40 keV and 1 sec time bin RXTE-PCA light curve of
IGR 17091-3624 on 04/08/2011. (b) 2 - 40 keV and 1
sec time bin RXTE-PCA light curve of $\rho$ class of GRS 1915+105.
Following \citet{b00} with 2-5 keV, 5-13 keV and 13-60 keV 
energy bands we draw (c) Color-color diagram of (a) in black cross 
with gray error-bars, color-color diagram of 
(b) in black stars with gray error-bars.
They look totally different. This indicates that the notion of soft and hard photons
may be totally different in these two objects implying that the masses should be different.
\label{fig4}}
\end{figure}

In order to show the mass independent aspect of CE, we first show that conventional
photon ratios do not look similar even when light curves of the same variability
class of the objects are chosen.
In Fig.~\ref{fig4}(a), we show the $2-40$ keV data of 04/08/2011 PCA observation of 
IGR 17091-3624 which look similar to the so-called $\rho$ class data of GRS 1915+105 
in Fig.~\ref{fig4}(b). In Fig.~\ref{fig4}(c), we draw the color-color diagram of GRS 1915+105 
(Fig.~\ref{fig4}b) following \citet{b00} with $2-5$keV, $5-13$keV, and $13-60$keV energy bands
with black crosses along with gray error bars. 
When we repeat the exercise for IGR 17091-3624 (Fig.~\ref{fig4}a) in Fig.~\ref{fig4}(c) with black 
stars and gray error bars, the ratio looks different, more like that of the 
$\phi$ class of \citep{b00}. With suitable adjustment
of energy ranges one can try to make them look similar, but this exercise is difficult to justify.
What is clear is that soft and hard photons with preassigned energy bands are not useful in
classifying variabilities. We need to take resort to a quantity which is independent 
of the mass of the black hole.

\begin{figure}
\includegraphics[angle=-90,scale=0.55]{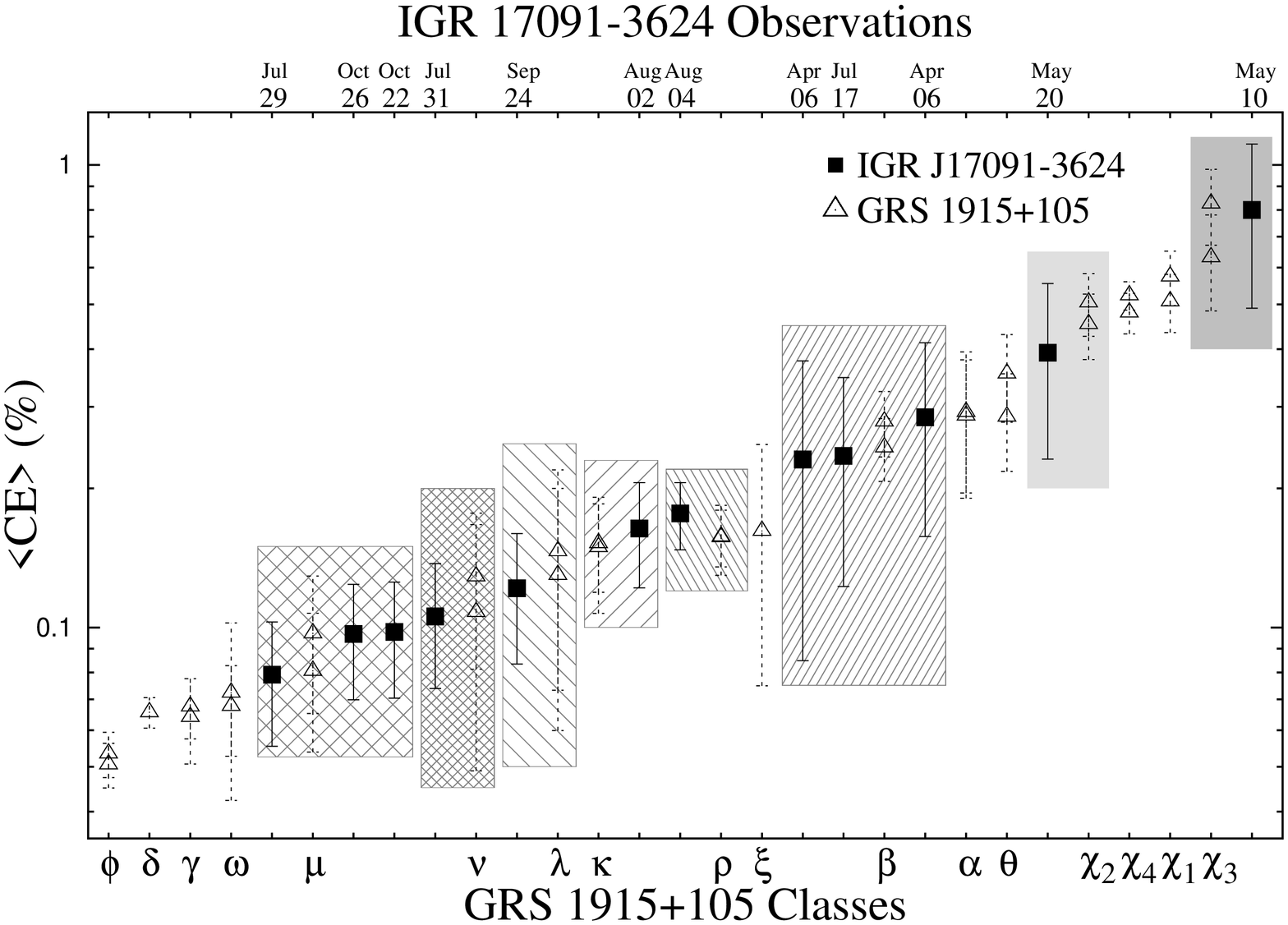}
\caption{Average CE of IGR 17091-2634 (filled square) for the
2011 outburst drawn in ascending order is compared with average CE of GRS 1915+105 (hollow triangles).
CE values of GRS 1915+105 (taken from the Fig.~4 of \citet{p13}). `Error bars' represent amount 
of excursion of 16s averaged CEs (including their 90\% confidence level error-bars) within these 
variability classes. Shaded boxes contain CE of GRS 1915+105 (two sets) and IGR 17091-3624 for 
which the light curves superficially are similar. Clearly, light curves of similar appearances 
produce roughly similar CEs.
\label{fig5}}
\end{figure}

In Fig.~\ref{fig5}, we plot average CE in each class for IGR 17091-3624 with filled squares. 
Average CEs are placed in increasing order. Dates of observations are given on the upper 
X-axis (see, Table.~\ref{tbl1}). For the sake of comparison, we plot average CEs of GRS 1915+105 with hollow 
triangles for the observed variability classes reported in \cite{c05, p11, p13}. 
Two sets of CEs of GRS 1915+105 (triangles) are taken from Fig.~4 of \citet{p13}. Shaded boxes group 
together light curves of 'similar appearance' from IGR 17091-3624 and GRS 1915+105 which were either 
reported by others or were discovered by us (see, Fig.~\ref{fig1}). 
Each average CE (filled square or triangle) has an `error bar' representing excursion of CE.
within that class. The upper limits come from the highest value of positive 90\% confidence error-bars
and the lower limits come from the highest value of negative 90\% confidence error-bars. 
See, \citet{p13} for details.

In Fig.~\ref{fig1}, we observe that the objects exhibit variabilities of 
similar appearance but the time scales are vastly different. For $\nu$, $\rho$ or $\kappa$ type
data, time-scale to complete an oscillation in IGR 17091-3624 is much shorter than that in GRS 1915+105.
As \citet{cm00} pointed out, the alternate softer and harder spectra in these intermediate
states could be  due to the interaction between the jets and the disk matter where the base of the jet
(the post-shock region) changes its optical depth periodically. This generally points to the fact that 
the black hole in IGR 17091-3624 could have much smaller mass than that in GRS 1915+105.
However, Fig. 5 shows that CEs of the two black hole candidates are very close to each other in all the 
common variability classes. This shows that perhaps the number of hard and soft 
photons also scaled in the same way as the mass of the black holes, but their ratio remains the same.
We have another reason to believe that the masses of IGR 17091-3624 and GRS 1915+105 are different. 
In GRS 1915+105, the variability classes were distinguished on the basis of 
color-color diagrams. However, in Fig.~\ref{fig4}c, we demonstrated that these diagrams drawn using
fixed energy bands could be misleading. They look completely different even when the light curves are
similar. This means that the so-called soft and hard photons do not belong to the same energy bands
in these two objects. This can be generally possible only if the masses are different. 
Note that this conclusion is solely based on spectral and timing features and the distances of 
these objects are immaterial. 
There is no possibility that the accretion rates of the disk and the halo can conspire to reproduce the
same spectral features at different timescales consistently for so many variability classes.
However, even then, CE values that we compute, are found to have similar values for both the objects
for the same class. Thus we believe that the evolution of the Comptonizing cloud in presence of 
seed photons of the standard disk, takes place exactly the same way in both the objects.

From our study we make some more observations:

(a) Several classes of light curves are not yet detected in IGR 17091-3624 (e.g., softer variability 
classes such as $\phi$, $\delta$, $\gamma$ and $\omega$). (b) No variability class transition
has been observed within a single dwell (unlike in GRS 1915+105 where this was seen in several IXAE
observations. See, \citet{c05}). We predict that given sufficient observation time, IGR 17091-3624
would also have similar CE values in yet unobserved variability classes and the {\it same sequence} in 
class transitions as that of GRS 1915+105 will prevail. On the other hand, given that the accretion rate
in IGR 17091-3624 is perhaps much lower \citep {al11c}, we suspect that it may not show the softer 
variability classes.  

Near equality of average CE in IGR 17091-3624 and GRS 1915+105 leads us to believe that average CE 
may characterize variability classes uniquely for any stellar mass black holes. The scaling behavior of the 
Compton cloud and the standard disk only establishes that the Compton cloud is an integral component 
of the accretion flow itself as in the two-component advection flow solution of CT95.  
The conventional color-color diagrams with fixed energy bands
cannot uniquely characterize the classes, since the
physical origin of photons of a given energy band is different when the mass changes.
This is not to say that the conventional color-color diagrams are not useful. 
Classifications of spectral states of neutron stars are successfully made on the basis of 
color-color diagrams from their energetic radiations \citep{has87, has89, sch89}. 
Since the masses of neutron stars fall within a narrow range, these diagrams drawn 
using {\it fixed} sets of energy bands, can be used to classify all the neutron stars easily. 
However, stellar mass black holes are known to have masses in the range of 
$3-20 M_\odot$ \citep{rm06} and this creates an obvious problem. 

Massive and super-massive black holes do have spectral states similar to stellar mass black holes.
Indeed, the two-component advective flow model was first postulated for the active galaxies
due to easy availability of low angular momentum matter in such systems \citet{c95}.
They also show short and long term variabilities (e.g., \citet{c12, e13} and references therein). 
It is not unlikely that the CE defined in our method will have similar 
values in these objects as well. This interesting aspect will be explored separately elsewhere.

\section{Acknowledgment}

PSP Acknowledges SNBNCBS-PDRA Fellowship.

\end{document}